\documentclass[journal,12pt,onecolumn,letterpaper]{IEEEtran}
\usepackage{arxiv}
\usepackage{geometry}
\usepackage{times}
\usepackage{cite}
\usepackage{url}
\usepackage{graphicx}
\graphicspath{{Images_SimplyMime/}}
\usepackage{lscape}
\usepackage{subfigure}
\usepackage{rotating}
\usepackage{rotfloat}
\usepackage{xcolor}
\usepackage{amsmath}
\usepackage{amssymb}
\usepackage[linesnumbered,ruled,vlined]{algorithm2e}
\usepackage{pseudocode}
\usepackage{array}
\usepackage[english]{babel}
\usepackage{gensymb}
\usepackage{textcomp}
\usepackage{placeins}
\usepackage{balance}
\usepackage{booktabs}

\title{The Qey: Implementation and performance study of post quantum cryptography in FIDO2 \\ 
}

\author{%
  
    Aditya Mitra \\
    Centre of Excellence, Artificial Intelligence \& Robotics (AIR),\\
    School of Computer Science and Engineering\\
    VIT-AP University, India \\
    DigitalFortress Private Limited \\
    \texttt{adityaarghya0@gmail.com}
\And

  Sibi Chakkaravarthy Sethuraman\\
    Centre of Excellence, Artificial Intelligence \& Robotics (AIR),\\
    School of Computer Science and Engineering\\
    VIT-AP University, India \\
    DigitalFortress Private Limited \\
    \texttt{sb.sibi@gmail.com} \\
}

\begin{document}
\maketitle

\begin{abstract}
Authentication systems have evolved a lot since the 1960s when Fernando Corbato first proposed the password-based authentication. In 2013, the FIDO Alliance proposed using secure hardware for authentication, thus marking a milestone in the passwordless authentication era \cite{lindermann2014securehardware}. Passwordless authentication with a possession-based factor often relied on hardware-backed cryptographic methods. FIDO2 being an amalgamation of the W3C Web Authentication and FIDO Alliance Client to Authenticator Protocol is an industry standard for secure passwordless authentication with rising adoption for the same \cite{fido2024passkeyadoption}. However, the current FIDO2 standards use ECDSA with SHA-256 (ES256), RSA with SHA-256 (RS256) and similar classical cryptographic signature algorithms. This makes it insecure against attacks involving large-scale quantum computers \cite{shor1999}. This study aims at exploring the usability of Module Lattice based Digital Signature Algorithm (ML-DSA), based on Crystals Dilithium as a post quantum cryptographic signature standard for FIDO2. The paper highlights the performance and security in comparison to keys with classical algorithms.
\end{abstract}

\section{Introduction}
 With the recent shift towards adoption of passkeys, it has been evident that cryptographic methods for authentication based on possession-based factors are more secure compared to knowledge-based factors. Passkeys are phishing resistant and easier to use compared to passwords \cite{ramat2025passkeyux} and is being adopted by more and more people and organizations to protect against sophisticated social engineering-based attacks \cite{fido2024passkeyadoption}. It is considered a state-of-the-art security standard. However, its reliance on classical cryptographic algorithms might be a major threat in the future when the attacks might involve large scale quantum computers.
Shor’s Algorithm is a quantum algorithm that threatens classical public key cryptography like ECC and RSA \cite{mishra2024quantum}. Shor’s algorithm is capable of calculating prime factorization and discrete logarithms on a quantum computer in polynomial time \cite{shor1999}, thus breaking the fundamental assumption of classical public key cryptography. Further, with the rise in quantum computers, it would be in near future when attackers, especially state-sponsored attackers would have large scale quantum computers under their controls.
FIDO2 works with W3C Web Authentication standard \cite{w3c2021webauthn2} and FIDO Alliance Client to Authenticator Protocol \cite{fido2021ctap}. It uses RS256 and ES256 algorithms by default and hence falling in the category of classical algorithms. Major bottlenecks towards Post Quantum Cryptography (PQC) adoption for FIDO2 includes:
\begin{itemize}
 \item Large key-sizes and signature sizes of Post quantum digital signature algorithms
 \item Unavailability of specialized hardware like Secure Elements (SE) or Trusted Platform Modules (TPM) for PQC
 \item Unavailability of a proper adoption roadmap.
\end{itemize}
This study aims to develop a physical security key following the post quantum cryptographic standards and the FIDO2 specifications so as to maintain compatibility with existing authentication standards and at the same time be secure against attacks involving large scale quantum computers.
The contributions in this paper:
\begin{itemize}
    \item First work to implement FIDO2 Authenticator with PQC
    \item Development on ARM based processor
    \item Benchmarking against classical algorithms implemented on similar hardware.
\end{itemize}

The rest of the document is as follows: Section \ref{topic1:l1} discusses the background and related work. Section \ref{topic1:l2}presents the proposed work. Section \ref{topic1:l3} discusses the design \& implementation. Finally, section \ref{topic1:l4} concludes our proposed work .

\section{Background and Related Work} \label{topic1:l1}
FIDO2 has been instrumental in modern passwordless authentication. A study by Klieme et al. \cite{klieme2020fidonous} has explored the usability of FIDO2 for continuous authentications on web services while another study \cite{angelogianni2024fidoprotocols} has pointed out that the existence of various different protocols like Universal 2-Factor (U2F), Universal Authentication Factor (UAF) and FIDO2 might be redundant. However, it is to be noted that these protocols are the natural evolution of a preliminary proof-of-concept to a widely adopted standard. U2F and UAF required special features of the web browser or extensions to work properly, but W3C later standardized the Web Authentication standard giving rise to the latest FIDO2. It makes FIDO2 universally running on all modern devices including any Android 7 or up, iOS 7 or up, Windows 10 version 1903 or up, Mac OS X or up and more. Our own study \cite{chakkaravarthy2022loki} has even showed the wide usability of FIDO2 in non-standard IoT devices and the usability to control authentication to physical assets.
However, it is to be noted that over the evolution of FIDO2, what has mostly evolved is the implementation while the underlying cryptographic primitives has stayed the same. ES256 (ECDSA with SHA-256) and RS256 (RSA with SHA-256) has been the default cryptographic algorithms used in FIDO2 for the past decade. While this has not caused any problem till-date, the rise of large-scale quantum computing shows that it is only a matter of time till these algorithms collapse. It is quite a good time to start migrating towards post quantum cryptography. However, the unavailability of a post quantum cryptographic accelerator has been a major bottleneck. 

A study \cite{mishra2024quantum} has shown that it is indeed possible to decompose classical cryptographic primitives with quantum computing on a quantum simulator. Shor’s Algorithm \cite{shor1999} is able to perform integer factorization and discrete logarithm in polynomial time on quantum circuits, which shatters the very assumptions of classical public-key cryptography that a computer needs exponential time for these operations. Another study \cite{shakib2025impersonation} has successfully shown this can be used to attack Blockchain-based networks, and hence is a matter of time till it attacks authentication systems.
Module Lattice based Digital Signature Standard (ML-DSA), as described in FIPS 204 \cite{nist2024mldsa} is a promising NIST PQC Finalist for Digital signature algorithm. This algorithm has some pros like being resistant to attacks involving large scale quantum computers and cons like large key and signature sizes. Hence the performance is subject to study on real hardware.

The proposed standard, ‘The Qey’ attempts to be the first to bring post quantum cryptography to FIDO2 based systems and performs a comparison with classical algorithms.

 \section{Proposed Work: The Qey} 
 \label{topic1:l2}
 
The Qey is a practical implementation based on the Internet Draft \cite{mitra2025mldsa} that gives a suggested procedure to bring ML-DSA to Web Authentication. The Qey is built on consumer grade hardware and is a proof of concept to testing ML-DSA for FIDO2. 
The system implements an ARM based microcontroller for processing ML-DSA due to the current unavailability of cryptographic accelerators. It uses a USB Controller, exposed as a USB-HID interface to leverage CTAP-HID like other security keys, without having to use any client-side application or additional drivers. The cryptographic secrets were stored in a specially permission-controlled way on an SD Card due to the unavailability of any secure storage media supporting PQC standards.

\subsection{Hardware Design overview of the Qey}
\begin{figure}[h!]
\centering
\includegraphics[width=0.8\linewidth]{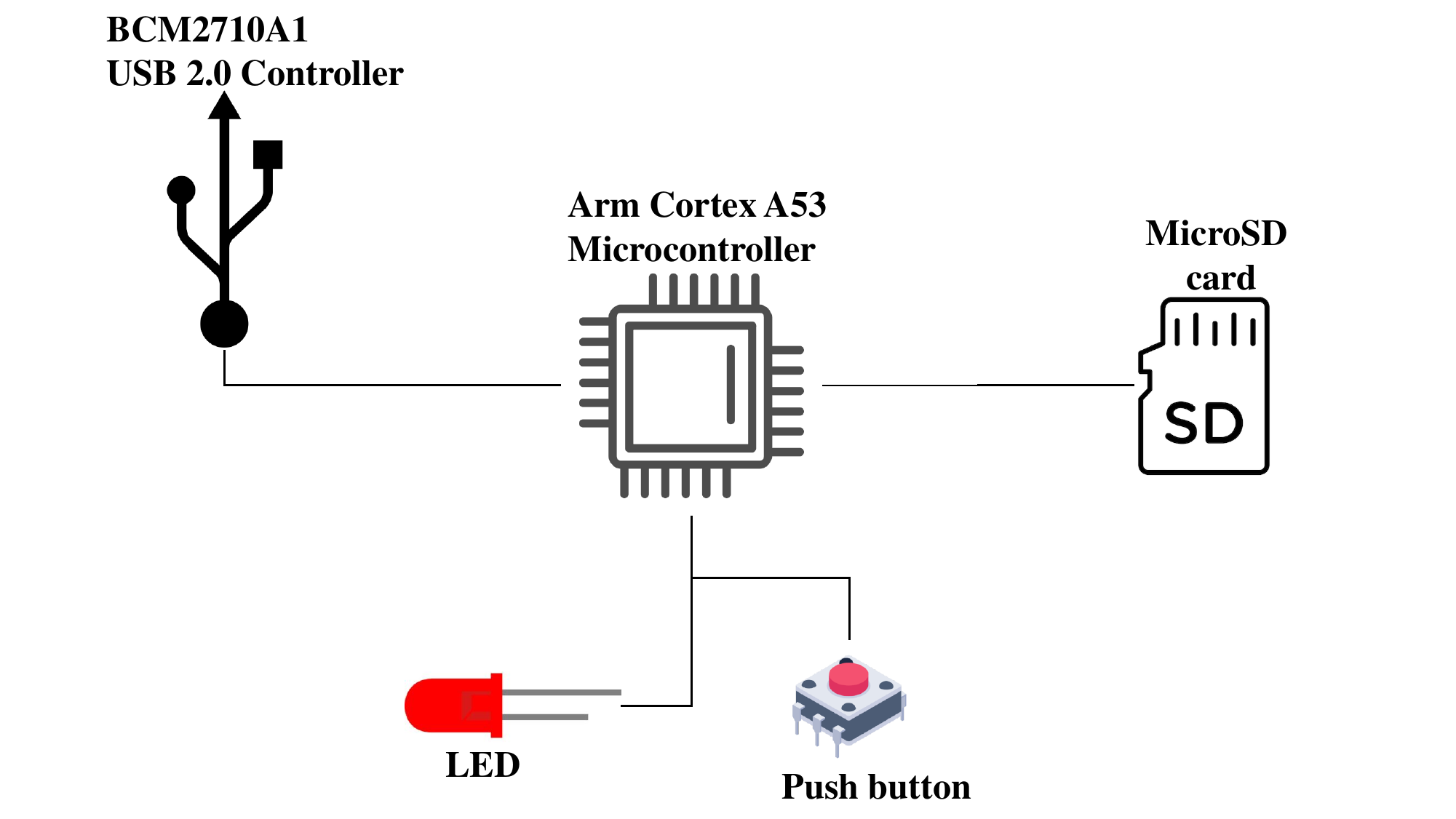}
\caption{The Qey}
\label{fig:Architecture}
\end{figure}

An ARM Cortex A-53 processor-based microcontroller is used as the processing unit for ML-DSA. The microcontroller is interfaced with a USB 2.0 port with a theoretical speed of upto 480 Mbps. A separate MicroSD card is used to hold the operating system and to hold the cryptographic secrets. A LED was added that could blink requesting user action and a push button was added for the User Presence checks. Figure \ref{fig:Architecture} shows the hardware configuration of the setup.

\subsection{Software design overview of the Qey}

The proposed system essentially works as a USB gadget. It ideally should have been developed as an embedded device but due to high computing power required for post quantum cryptography, this was not feasible. Hence, the system is designed to run on a stripped-down Debian based operating system. A python-based implementation of CTAP was written from scratch and was tested to work with standard FIDO2 services. The application exposes the USB port as a HID device with a device descriptor having Usage Class 0xD0, 0xF1. This signifies to the host computer, when connected, to use this device as a FIDO key. 
The device can read/write HID data from the USB as a file /dev/hidg0. The device uses a ML-DSA functions from Open Quantum Safe (OQS), which has been recommended by NIST as an implementation of post quantum cryptographic algorithms \cite{baentsch2021oqs} - \cite{stebila2017oqs}. 
FIDO2 relies on COSE assignments for the cryptographic algorithms. However, IANA has not yet assigned any number to ML-DSA algorithms. However, the Cose WG at IETF recommends to adopt the number -48, -49, -50 for ML-DSA with parameter set 44, 65, and 87 respectively in an Internet Draft \cite{prorock2025mldsa}. For the current prototype of The Qey, it has been assumed that these assignments will be accepted by IANA in future.
The python-based implementation of CTAP incorporating ML-DSA is set to run as a systemd service on the device, thus working as soon as the device is plugged in to a host.

\subsection{CTAP Implementation for The Qey}
The proposed standard follows a request-response communication protocol over USB HID as per the Client to Authenticator Protocol (CTAP). It implements all major functionalities required for successful registration and authentication.  Table \ref{Tab1. Implementation differences} shows the implementations in Qey and its deviation from usual implementations.

\begin{table}[h!]
\centering
\caption{Implementation differences with standard}
  \label{Tab1. Implementation differences}
\begin{tabular}{|p{6cm}|p{5.5cm}|p{5.5cm}|}
\hline
\textbf{Authenticator API Command} & \textbf{Implementation in The Qey} & \textbf{Standard Implementations} \\ \hline
\texttt{authenticatorClientPin (0x06)} & Supports Pin Protocol 1 only. & Supports Pin Protocols 1 and 2 \\ \hline
\texttt{authenticatorGetInfo (0x04)} & Returns Pin Protocol 1 as the only supported protocol for field \texttt{pinProtocols (0x06)} & Supports both protocols 1 and 2 as supported protocols \\ \hline
\texttt{authenticatorMakeCredentials (0x01)} & Supports algorithms \texttt{-48} and \texttt{-49} for \texttt{pubKeyCredParams (0x04)} field. & Does not support algorithms \texttt{-48} and \texttt{-49}. It supports only classical algorithms like \texttt{-7 (ES256)} and \texttt{-257 (RS256)} \\ \hline
\texttt{authenticatorGetAssertion (0x02)} & Can return ML-DSA signatures. & Can return only classical signatures. \\ \hline
\texttt{authenticatorGetNextAssertion (0x08)} & Can return ML-DSA signatures. & Can return only classical signatures. \\ \hline
\end{tabular}
\end{table}

The Qey supports the algorithms ML-DSA-44, ML-DSA-65 for Post Quantum and ES256 as a legacy fallback for backward compatibility of Relying parties that do not support ML-DSA yet. Table \ref{Tab2.Algorithms in the Qey} shows the algorithms.

\begin{table}[h!]
\centering
\caption{Algorithms in The Qey}
\label{Tab2.Algorithms in the Qey}
\begin{tabular}{|l|l|l|l|l|}
\hline
\textbf{Algorithm Name} & \textbf{Private Key Size} & \textbf{Public Key Size} & \textbf{Signature Size} & \textbf{Remarks} \\ \hline
\texttt{ML-DSA-44} & 2560 bytes & 1312 bytes & 2420 bytes & Suggested algorithm \\ \hline
\texttt{ML-DSA-65} & 4032 bytes & 1952 bytes & 2560 bytes & Slower algorithm \\ \hline
\texttt{ES256} & 32 bytes & 64 bytes & 64 bytes & Classical fallback \\ \hline
\end{tabular}
\end{table}

\section{Experimental Setup} 
\label{topic1:l3}

The key was built using a Raspberry Pi Zero 2W microcontroller. It is built on the ARM Cortex A-53 Quad Core CPU. It has 512 MB of LPDDR2 RAM which was enough to run the required functionalities. It also contains a Broadcom BCM2710A1 USB 2.0 Controller which became fundamental for it to act as the USB HID device. A 16GB MicroSD card was used to hold the operating systems and required program files. The functionality was developed in Python entirely, using Open Quantum Safe (OQS) for the ML-DSA cryptographic primitives and the Cryptography library for the classical fallback of ES256. A USB-A addon board was used to give it the form-factor of a USB stick, similar to any other physical security key. It drew an Idle power of 0.5w, making it easy to power by directly plugging it into the host computer. It did not need any external power source.

\begin{figure}[h!]
\centering
\includegraphics[width=0.8\linewidth]{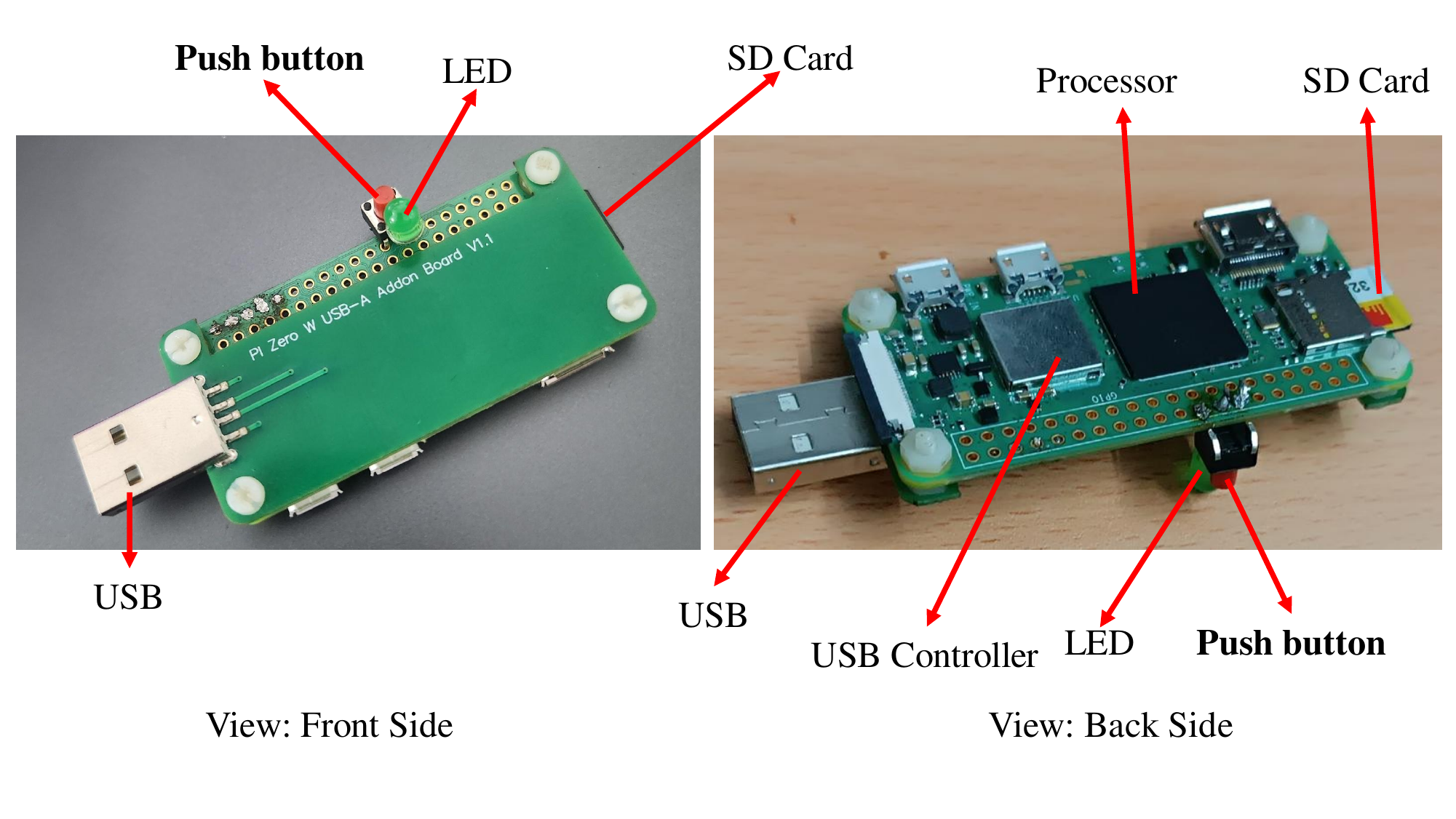}
\caption{The Qey}
\label{fig:TheQey}
\end{figure}

A LED and a pushbutton were soldered on the GPIO pins which could act as an indicator and a trigger respectively. The overall dimensions of the device come out to 65mm in length, 30mm in width and 12.2 mm in height (majorly due to the standard LED, which can be reduced by using a flat LED). Figure \ref{fig:TheQey} shows a prototype of The Qey.

\subsection{Performance analysis}
The Qey was tested with both classical algorithms and post quantum algorithms. Theoretically, the key and signature sizes of ML-DSA are way longer than that of ECDSA and hence Post Quantum Algorithms should have taken considerably more time than classical algorithms. However, on performance analysis with over 30 samples, it was discovered that the ML-DSA was slower on the order of 10000 micro seconds or by 10 milliseconds only. This is well within the acceptable range of time taken for authentication and hence proves that with suitable CPUs, it is quite possible to make the computation time for PQC similar to that of classical computing. Table. \ref{Tbl3:Average Time Per Operation} shows the average time taken per computation.

\begin{table}[h!]
\centering
\caption{Average Time Taken per Operation}
\label{Tbl3:Average Time Per Operation}
\begin{tabular}{|l|l|l|}
\hline
\textbf{Function} & \textbf{Algorithm} & \textbf{Average Time (Micro-sec)} \\ \hline
Authentication & \texttt{ES-256} & 3192.7 \\ \hline
Authentication & \texttt{ML-DSA-44} & 17800.6 \\ \hline
Authentication & \texttt{ML-DSA-65} & 30675.2 \\ \hline
Registration & \texttt{ES-256} & 12251.8 \\ \hline
Registration & \texttt{ML-DSA-44} & 36069.2 \\ \hline
Registration & \texttt{ML-DSA-65} & 68086.8 \\ \hline
\end{tabular}
\end{table}

Authentication time was measured as the total time to respond to a ‘authenticatorGetAssertion’ call which includes the time taken for reading the command, retrieving the cryptographic secret from the MicroSD card, signing the challenge, fragmenting the output into frames of 64 bytes, and sending the output to the host. Similarly, the Registration time was measured as the total time to respond to a ‘authenticatorMakeCredential’ call which includes the time taken for reading the command, generating keypair, writing the cryptographic secret to the MicroSD card, generating attestation, fragmenting the output to 64-byte frames and sending it over to the host. 
The host computer was entirely unmodified, running Windows 11 with Microsoft Edge browser, using the Windows default Web Authentication handler from Windows Hello. The relying party server was developed as a fork from a standard library by Yubico with ML-DSA signature verification added with OQS. Hence, it is evident that the host machine is able to support ML-DSA based Web Authentication without any changes.

\subsection{Attack Analysis}
An attack analysis of the standard shows that by following all FIDO2 standards, the Qey is resilient to most common attacks like Phishing and MITM.
\begin{itemize}
\item Phishing: FIDO2 specifications enforce verification of Relying Party identity before any cryptographic operation. The usage of TLS verifies the legitimacy of the website and protects against DNS based attacks.
\item Man-in-the-middle (MITM): Web Authentication works only over TLS which ensures all communication between the relying party server to the host is end to end encrypted, mitigating any risk of MITM.
\item Harvest now, decrypt later (HNDL): The use of post quantum algorithms ensure that the secrets cannot be discovered by an unauthorized entity even if the communications are captured now and the entity waits for a breakthrough in decryption technologies with a Quantum computer.
\end{itemize}

\section{Conclusion} \label{topic1:l4}
This paper presents a standard that uses post quantum cryptography for FIDO2 based authentication on websites, applications and services. It maintains all security principles and phishing-resistance of FIDO2 while upgrading the cryptography to be resilient to attacks involving a large-scale quantum computer. This enhances the security of the device against HNDL attacks, which current keys are vulnerable to. However, due to the current lack of technology, the Qey is vulnerable to attacks whether the attacker may physically get hands on the key. This is because of the lack of secure storage devices like SE and TPMs supporting post quantum cryptography. This will be patched soon with hybrid cryptographic approaches.
Further, in future versions of The Qey, it is aimed to use biometrics instead of the pin with Pin protocol 1. This is because Pin Protocols 1 and 2 both rely on classical cryptography and would be unusual for a post quantum security key.


\section*{Acknowledgments}
This research was conducted as part of the Research and Development (R\&D) efforts at DigitalFortress Private Limited \& Indominus Labs Private Limited. We acknowledge the company's support in facilitating this study. This work is protected and is not intended for reuse in any commercial capacity.

\bibliographystyle{IEEEtran}  
\bibliography{references}  

\end{document}